\documentclass[epj]{webofc}
\usepackage[varg]{txfonts}   

\usepackage{graphicx,color} 
\usepackage{bm}       
\usepackage{amsmath}  
\usepackage{amssymb}
\usepackage{color}
\usepackage[utf8]{inputenc}

\newcommand{\feynslash}[1]{\slash\!\!\! #1}
\newcommand{\traceD}{\mathrm{tr}_\mathrm{CD}\!\ }
\woctitle{21st International Conference on Few-Body Problems in Physics}
\begin{document}
\title{Contemporary continuum QCD approaches to excited hadrons}
\author{Bruno El-Bennich\inst{1,2}\fnsep\thanks{\email{bruno.bennich@cruzeirodosul.edu.br}} \and
            Eduardo Rojas \inst{3}\fnsep\thanks{\email{eduardo.rojas@cruzeirodosul.edu.br}} 
          } 

\institute{Laborat\'orio de F\'isica Te\'orica e Computacional, Universidade Cruzeiro do Sul, Rua Galv\~ao Bueno 868, 01506-000 S\~ao Paulo, SP, Brazil
\and
  Instituto de F\'isica Te\'orica, Universidade Estadual Paulista, Rua Bento Teobaldo Ferraz 271, 01140-070 S\~ao Paulo, SP, Brazil
\and
  Instituto de F\'isica, Universidad de Antioquia, Calle 70, no.  52-21, Medell\'in, Colombia
  }

\abstract{Amongst the bound states produced by the strong interaction, radially excited meson and nucleon states offer an important phenomenological 
window into the long-range behavior of the coupling constant in Quantum Chromodynamics. We here report on some technical details related to
the computation of the bound state's eigenvalue spectrum in the framework of Bethe-Salpeter and Faddeev equations.
}
\maketitle

\section{Introduction: excited states as an eigenvalue problem}
\label{sec1}

A great deal of research activity in hadron physics is concerned with hadron structure and revolves around two fundamental questions:
what constituents are the hadrons made of and how does Quantum Chromodynamics (QCD), the strong interaction component of the Standard
Model, produce them? These are simple questions which, however, may not entail simple answers. To understand the measurable content
of QCD, spectroscopy is a valuable and time-honored tool --- suffice it to mention the inestimable progress made in the computation of atomic or 
molecular spectra and subsequent comparison with experiments that lead to a deeper understanding of Quantum Electrodynamics.

The same is true for QCD: if confinement is related to the analytic properties of QCD's Schwinger functions, then light-quark confinement ought to 
be understood  by mapping out the infrared behavior of the theory's universal $\beta$ function. Obviously, this cannot be possibly achieved in perturbation
theory. A nonperturbative continuum approach to QCD is provided by Dyson-Schwinger equations (DSE) which relate the theory's $\beta$ function 
to experimental observables~\cite{Bashir:2012fs}.  Therefore, comparison between DSE predictions embedded in bound-state calculations and 
observations of the hadron mass spectrum as well as of elastic and transition form factors can be used to study the long-range behavior of QCD's 
interaction. As the properties of excited hadron states are considerably more sensitive  to the long-range behavior of the strong interaction than those of 
ground states~\cite{Qin:2011xq,Rojas:2013tza,Rojas:2014aka}, excited mesons and  nucleons~\cite{Aznauryan:2012ba} are an important source of information 
and complement our understanding of the strong interaction in light mesons~\cite{ElBennich:2008qa,Chang:2009ae,ElBennich:2012ij,daSilva:2012gf} 
and in heavy-light mesons with disparate energy scales~\cite{ElBennich:2009vx,ElBennich:2010ha,ElBennich:2011py,ElBennich:2012tp}.

Of course, the properties of these hadrons can only be understood, and the functional behavior of the $\beta$ function be inferred therefrom, 
by studying the quark's DSE in conjunction with quark-antiquark or three-quark bound-state equations, the Bethe-Salpeter and Faddeev equations,
respectively. Both are treated as eigenvalue problems which we here exemplify with the relativistic bound-state equation of pseudoscalar $J^P= 0^-$ 
mesons.  The homogeneous Bethe-Salpeter equation for this $q\bar q$ bound state with relative momentum $p$ and total momentum 
$P$ can be generally written as,
\begin{equation}
\label{BSE} 
  \Gamma_{0^-} (p,P) = \int^\Lambda_k\!\!  \mathcal{K} (p,k,P)\left  [ S (k + \eta_+P)\, \Gamma_{0^-}  (k,P)\, S (k-\eta_- P) \right ]  \ ,
\end{equation}
where $S (k\pm \eta_\pm P)$ are dressed quark propagators with $\eta_+ +\eta_- =1$ (NB: in a Poincar\'e invariant calculation numerical 
results are independent of the momentum partition parameter $\eta_\pm$). For the sake of simplicity, we omit flavor and Dirac indices, as the 
following discussion is independent of them. In the same spirit, we restrict ourselves to the rainbow-ladder truncation of the interaction kernel,
\begin{equation}
\label{BSEkernel} 
   \mathcal{K} (p,k,P) =  - \frac{Z_2^2\, \mathcal{G} (q^2 )}{q^2} \, \frac{\lambda^a}{2}  \gamma_\mu  T_{\mu\nu}(q) \,
    \frac{\lambda^a}{2}  \gamma_\nu  \  ,
\end{equation}
with the transverse projection operator $T_{\mu\nu} (q) :=  g_{\mu\nu} -  q_\mu q_\nu/q^2 $, $q=p-k$, $Z_2$ is the wave-function renormalization
constant and $\lambda^a$ are the SU(3) color matrices in the fundamental representation. Various model ans\"atze~\cite{Maris:1997hd,
Maris:1997tm,Maris:1999nt,Qin:2011dd} have been proposed for the effective interaction, $\mathcal{G} (q^2 )$, which emulates
the combined effect of the gluon and quark-gluon vertex dressing functions and most recent efforts extend these models to include 
important transverse components of this vertex beyond the leading truncation~\cite{Binosi:2014aea}. Most importantly, 
Eq.~(\ref{BSEkernel}) satisfies the axialvector Ward-Takahashi identity~\cite{Bender:1996bb} and therefore ensures a massless pion in the 
chiral limit. As we shall see below, Eqs.~(\ref{BSE})  and (\ref{BSEkernel}) define an eigenvalue problem with physical solutions at the 
mass-shell points, $P^2 = -m^2$, where $m$ is the bound-state mass. 

The Bethe-Salpeter equation's Poincar\'e-invariant solutions can be cast in the form,
\begin{equation}    
  \Gamma_{0^-} \, (p, P)  =  \gamma_5\left [\, i\, \mathbb{I}_D  E_{0^-}  (p,P) +  \feynslash P F_{0^-}  (p,P)
                +\ \feynslash p (p\cdot P)\, G_{0^-}  (p,P) + \sigma_{\mu\nu}\, p_\mu P_\nu\,   H_{0^-}  (p,  P) \, \right ]  \ , 
\label{diracbase}                            
\end{equation}
Note that this Euclidean-metric basis, $\mathcal{A}^\alpha (p,P) =  \gamma_5 \big \{ i\,\mathbb{I}_D$,  
$\feynslash P$,  ${\feynslash p} (p\cdot  P)$, $\sigma_{\mu\nu}p_\mu P_\nu \big \}$, is nonorthogonal with respect to the Dirac trace. The 
functions $\mathcal{F}_{0^-} ^\alpha (p,P) =$  $\big \{ E_{0^-}  (p,P), F_{0^-}  (p,P), G_{0^-}  (p,P), H_{0^-}  (p,P)\big \}$ are Lorentz-invariant 
scalar amplitudes and are extracted from the Bethe-Salpeter amplitude~(\ref{diracbase}) with appropriate projectors, 
$P^a(p,P)$,\footnote{\ see, e.g., Ref.~\cite{Cobos-Martinez:2010raa} for the derivation of the change of basis coefficients $P^{\alpha\beta}(p,P)$.}
\begin{equation}
    \tfrac{1}{4}\, P^{\alpha\beta}(p,P)\, \mathrm{Tr}_\mathrm{D}
    \left [ \mathcal{A}^\beta (p,P)\mathcal{A}^\gamma (p,P)\right ]   = \delta_{\alpha\gamma} \ ; \quad 
    P^\alpha (p,P)  = P^{\alpha\beta}(p,P)\, \mathcal{A}^\beta (p,P) \  ,
\end{equation}
where $\alpha,\beta =1,...,4$ and the projection is given by:
\begin{equation} \hspace*{-1mm}
   \mathcal{F}_{0^-} ^\alpha (p,P) =  \tfrac{1}{4}\,  \mathrm{Tr}_\mathrm{D} \Big [ P^\alpha(p,P)\,\Gamma_{0^-}  (p,P)  \Big ]   \ .
  \label{project1}
\end{equation}
Using the Bethe-Salpeter equation, this leads to the eigenvalue problem, 
\begin{equation}
  \lambda(P^2)\, \mathcal{F}_{0^-} ^\alpha (p,P) =\int^\Lambda_k \!\! \mathcal{K}^{\alpha\beta} (p,k,P)\, \mathcal{F}_{0^-} ^\beta (k,P) \ ,
 \label{eigenvalue}
\end{equation}
where  $\mathcal{K}^{\alpha\beta} (p,k,P)$ stems from the projection of Eq.~(\ref{BSE}) using Eqs.~(\ref{BSEkernel}) and (\ref{project1}):
\begin{equation}
  \mathcal{K}^{\alpha\beta} (p,k,P) = -\frac{Z_2^2}{4} \frac{\mathcal{G} (q^2 )}{q^2}\, T_{\mu\nu}(q)\,  \traceD \Big [ P^\alpha (p,P) \,
   \gamma_\mu\, \lambda^a  S (k_+)\,  \mathcal{A}^\beta (k,P)\, S (k_-)\,  \gamma_\nu\, \lambda^a  \Big ] \ .
\label{KernelAB}
\end{equation}
In Eq.~(\ref{eigenvalue}), $\lambda(P^2)$ is a scalar function and the eigenvalue equation has a solution for every value of $P^2$. 
Typically, iterative eigenvalue algorithms are employed as only under simplifying assumptions an inversion of the Bethe-Salpeter problem 
is possible~\cite{Lucha:2012ky,Frederico:2013vga,Bhagwat:2007rj}. To elucidate the iterative procedure, we simplify Eq.~(\ref{eigenvalue}) 
(see also Ref.~\cite{Krassnigg:2003wy}), so that:
\begin{equation}
    \lambda(P^2)\, | \Phi \rangle = \mathcal{K} (P^2)\, | \Phi \rangle \ . 
\end{equation}
The kernel, $\mathcal{K} (P^2)$, has a complete set of real eigenvectors $\phi_i$ with eigenvalues $\lambda_i (P^2)$ which are 
ordered as $\lambda_0 (P^2) > \lambda_1 (P^2) > \lambda_2 (P^2) > .... > \lambda_i (P^2)$. Thus, any solution can be written as
a linear superposition,
\begin{equation}
  | \Phi \rangle = \sum_{i=1}^\infty a_i\, |\, \phi_i \rangle \ ,
  \label{guess}
\end{equation}
where $a_i$ are real constants and the vector must be normalized.  To begin the iterative process, one may ``guess'' as solution, 
such as in Eq.~(\ref{guess}), for a given value of $P^2$ and $n$ successive actions of the kernel lead to, 
\begin{equation}
  |\, \phi_n \rangle := \, \mathcal{K}^n (P^2)\, | \Phi \rangle = \sum_{i=1}^\infty\, \lambda^n_i  a_i\, |\, \phi_i \rangle = 
   \lambda_0^n \left [ a_0\, |\, \phi_0 \rangle + \sum_{i=1}^\infty \left ( \frac{\lambda_i}{\lambda_0} \right )^n\! a_i\, |\, \phi_i \rangle \right ] \ .
\end{equation}
Since $\lambda_0 > \lambda_i$, the coefficients of $|\, \phi_i \rangle$ converge to zero for sufficiently large values of $n$ and therefore
the amplitude $|\, \phi_n \rangle$ converges to the ground state amplitude $|\, \phi_0 \rangle$:
\begin{equation}
  |\,\phi_n \rangle \overset{n\to \infty}{=}  \lambda_0^n\,  a_0\, |\, \phi_0 \rangle\, \simeq \, \lambda_0\, \mathcal{K}^{n-1} (P^2)\, | \Phi \rangle \ .
\end{equation}
This is the most basic method of computing the largest eigenvalue $\lambda_0(P^2)$ and its associated eigenvector to any required accuracy 
and is referred to as power or {\em von Mises\/} iteration. The trajectory of this eigenvalue function for a range of $P^2$ values yields the ground-state
meson mass, that is one finds $\lambda_0(P^2) =1$ when $P^2= -m_0^2$.

\section{Gram-Schmidt orthogonalization}
\label{sec2}

\begin{figure}[t!]
\vspace*{-2mm}
\centering
\includegraphics[width=0.5\textwidth]{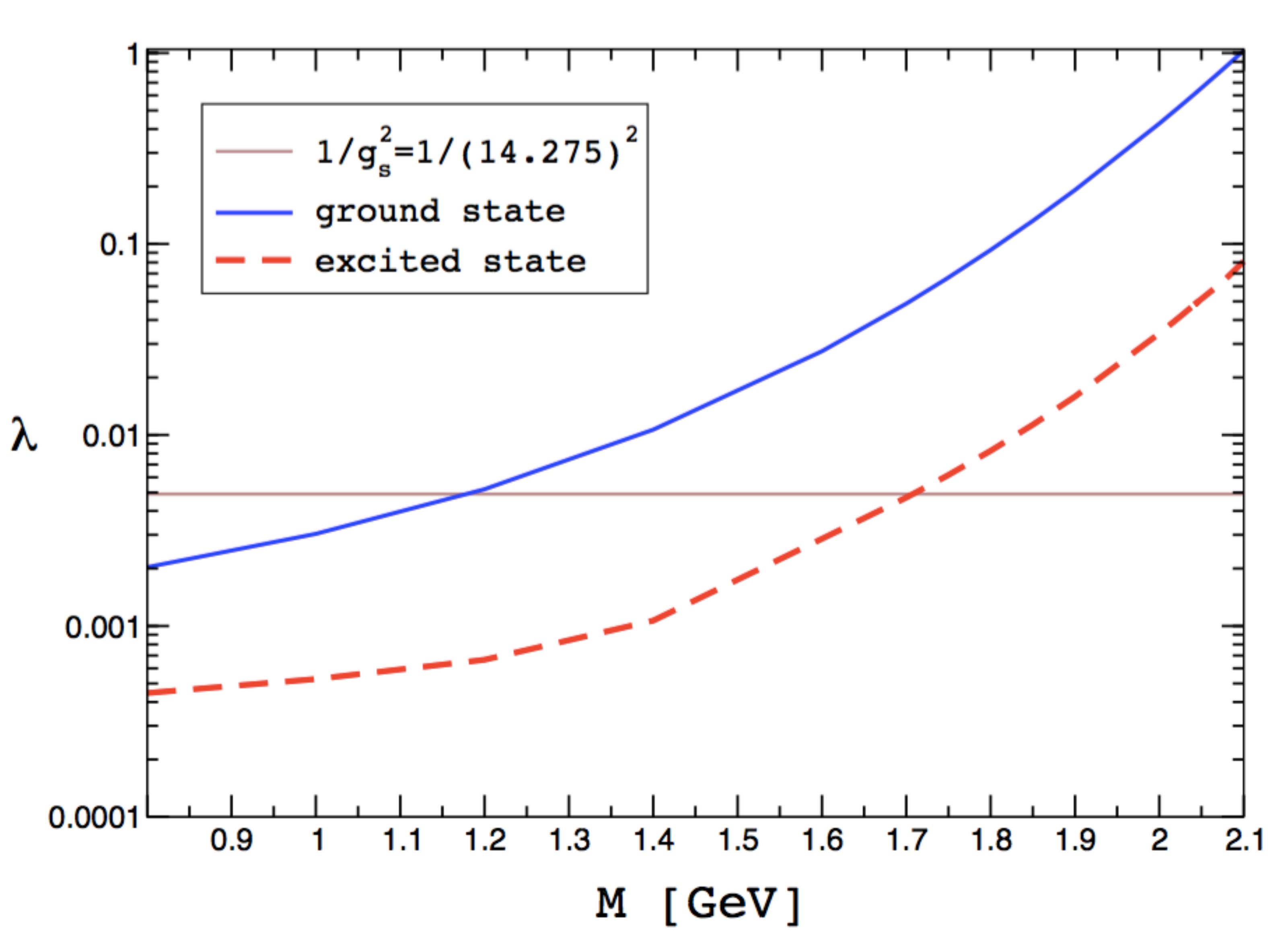}
\caption{First two eigenvalue trajectories of the quark-diquark kernel used in Refs.~\cite{Cloet:2008re,Segovia:2015hra} as a function 
              of $\sqrt{-P^2}$. Note that the eigenvector solutions are here differently normalized so that the eigenvalues are $\lambda_i=1/g_s^2$
              instead of $\lambda_i =1$ for $P^2=-m_i^2$; see Ref.~\cite{Cloet:2008re} for details. The intersections of the solid horizontal 
              line with the trajectories locates the mass-pole position of the nucleon and its first excited state identified with the {\em Roper\/}.}
\label{fig1}
\vspace*{-4mm}
\end{figure}

A common procedure to extract the wave function of excited states from the spectrum of the interaction kernel is based on
the Gram-Schmidt orthogonalization process. As mentioned in Section~\ref{sec1}, the homogeneous Bethe-Salpeter equation 
is an eigenvalue problem, where the largest eigenvalue the interaction kernel produces corresponds to the ground state. 
However, the eigenvalue spectrum is not limited to the ground state and excited states with smaller eigenvalues can be determined
with the same iterative methods discussed above. As sketched, e.g. in Ref.~\cite{Krassnigg:2003wy}, one chooses a mass that
is larger than the ground state mass,  $P^2 = -m^2 < -m_0^2$, and applies the iterative procedure outlined in Section~\ref{sec1}: 
\begin{itemize}
\item find the  largest eigenvalue, $\lambda_0$, and the associated eigenvector for $m^2 > m_0^2$ and
$\lambda_0(m^2) > \lambda_0(m_0^2)$. This is the unphysical ``ground state'' at the mass scale $P^2=-m^2$. 
\item make again a guess for the Bethe-Salpeter amplitude {\em now projecting out\/} the eigenvector that pertains to the 
largest eigenvalue $\lambda_0 (m^2)$ in a first iteration.
\item use the eigenvector obtained in the previous step as input for the Bethe-Salpeter amplitude and project out {\em again\/} 
the eigenvector associated with $\lambda_0(m^2)$ in a second iteration; the resulting eigenvector must be projected as before in a 
third  iteration and so on. 
\item the iteration converges after the $n$th projection which yields an eigenvalue, $\lambda_1(m^2)$, with an eigenvector orthogonal 
to the one associated with the ``ground state'' eigenvalue $\lambda_0(m^2) > \lambda_1(m^2)$.
\item  by varying $P^2$ one obtains the mass evolution of the second largest eigenvalue $\lambda_1(P^2)$, which is exemplified in
Figure~\ref{fig1}. The solution for $\lambda_1(P^2) = 1$ yields the mass of the first excited state, $P^2=-m_1^2$ where $m_1 >  m_0$, 
at which  the eigenvector  is the state's Bethe-Salpeter amplitude.  
\end{itemize}

\noindent
The normalized projection within this Gram-Schmidt procedure is effected by, 
\begin{equation}
  | \tilde \Phi \rangle =  | \Phi \rangle -   \frac{\langle \phi_0\, |\, \Phi \rangle}{\langle \phi_0\, | \, \phi_0 \rangle} \ | \phi_0 \rangle \ ,
 \label{project}
\end{equation}
where $|\Phi \rangle$ is the initial guess for the Bethe-Salpeter amplitude in Eq.~(\ref{diracbase}) and $|\, \phi_0 \rangle$ is the vector of 
the ground-state Bethe-Salpeter amplitude. In projecting out this ground-state, one must define a norm in Euclidean space via an inner 
product, as is evident from Eq.~(\ref{project}). Within Quantum Field Theory in rainbow-ladder approximation, this product is defined as:
\begin{equation}
   \langle \Psi\, |\, \Phi \rangle  := \traceD \!  \int_k^{\Lambda} \! \overline \Psi (k,-P)\, S(k+\eta_+ P)\, \Phi(k,p)\, S(k-\eta_- P) \ .
   \label{inner}
\end{equation}
Orthogonality is thus defined by the condition, $\langle \Psi | \Phi \rangle = 0$, which expresses the vanishing of the overlap amplitude 
at $P^2$. Note that Eq.~(\ref{inner}) is not valid for computations beyond the leading approximation, i.e. the rainbow-ladder truncation.

\section{Krylov subspace and Arnoldi iteration}
\label{sec3}

After momentum discretization and Chebyshev expansion of Eq.~(\ref{eigenvalue}), the numerical kernel of the Bethe-Salpeter equation
(or the Faddeev equation for the three-body problem) is a non-symmetric matrix of large dimensions and with some of its eigenvalues 
close to zero.  In general, non-symmetric matrices have eigenvalues and eigenvectors which are very sensitive to small changes in the 
matrix elements due to the lack of symmetries on which traditional methods rely to ensure numerical stability~\cite{Press:1992zz}. 
This is particularly true for radial excitations, where angular dependence encoded in higher Chebyshev moments contributes in a nontrivial 
manner. Therefore, we make use of the numerical  \texttt{ARPACK} library~\cite{Arpack} which is designed to compute a few eigenvalues and 
corresponding eigenvectors of a general $n\times n$ matrix by means of the implicitly restarted Arnoldi method (IRAM).
The strength of the Arnoldi method~\cite{Arnoldi:1951}  lies in the application of the stabilized orthogonalization algorithm in the Krylov space of a 
given matrix $K$ to find its eigenvectors, as will be explained shortly. The most important feature of this algorithm is its ability to decouple the 
calculation of the eigenvectors  corresponding to the eigenvalues with the largest absolute value from the eigenvectors whose eigenvalues have 
an absolute value close to zero. This is highly desirable as the calculation of the latter are plagued by numerical instabilities.

In algebra, the order-$r$ Krylov subspace is generated by an operator whose representation is given by an $n\times n$ matrix $K = K_{ij}$, 
and a vector $\Phi$ of dimension $n$. It corresponds to the linear subspace spanned by actions of $K$ on $\Phi$:
\begin{equation}
   \mathcal{S}_r\, := \, \left \{ \Phi, K \Phi, K^2 \Phi, K^3\Phi , .... , K^{r-1} \Phi \right \} \ .
\label{krylov}
\end{equation}
The power iteration of $K$ yields the sequence in Eq.~(\ref{krylov}) and converges to the eigenvector associated with the largest 
eigenvalue $\lambda_0$. However, hereby one only makes very limited use of the stored information, as only the final vector, $K^{r-1}\Phi$, is kept. 
On the other hand, the basis for the Krylov subspace is derived from the Cayley-Hamilton theorem which implies that the inverse of a matrix can be 
expressed as a linear combination of its powers. Thus, Krylov subspaces play an important role in contemporary iterative methods 
to obtain one or few eigenvalues of large sparse matrices or to solve large systems of linear equations. Instead of heavy matrix
operations, these methods rely on successive multiplications of vectors by the matrix, thus forming a Krylov subspace, and then employ 
the resulting vectors. 

The vectors of the Krylov space are initially not ortogonal and usually become almost linearly dependent due to the properties 
of the matrix power iteration. Nonetheless, an orthogonal matrix can be constructed from the basis vectors by means of the Gram-Schmidt 
process described in Section~\ref{sec2}. This method proves to be unstable but its shortcoming can be overcome with the Arnoldi 
iteration~\cite{Arnoldi:1951} which uses the {\em stabilized\/} Gram-Schmidt process and can be applied to general, possibly non-Hermitian
matrices. The Arnoldi method  generalizes the Gram-Schmidt process by computing the eigenvalues of the orthogonal projection of $K$ 
onto the  Krylov subspace, where the projection is represented by the upper Hessenberg matrix $H_r$~\cite{Arpack}. For Hermitian (symmetric) 
matrices, the Arnoldi iteration is analogous to the Lanczos iteration.

\section{A note on orthogonality}
\label{sec4}

In the context of Bethe-Salpeter and Faddeev equations, orthogonality is defined by Eq.~(\ref{inner}). In vectorial form, using the shorthand 
$x=k^2$ and $y=p^2$, Eq.~(\ref{eigenvalue}) can be written as,
\begin{equation}
   \lambda (y)\, \bm{F}_R(y)  =  \int\! dx\  \bm{\mathcal{K}}(x,y)\cdot \bm{F}_R(x) \ ,
   \label{eigenvalueeq}
\end{equation}
which in a numerical treatment is evaluated as the sum,
\begin{equation}
   \lambda(x_j) \, \bm{F}_R(x_j)  =  \sum_i w_i \, \bm{\mathcal{K}}(x_j,x_i) \cdot \bm{F}_R(x_i) \ ,
  \label{sum}
\end{equation}  
and similarly,
\begin{equation}
   \lambda(x_j) \, \bm{F}_L(x_j)  =  \sum_i w_i \, \bm{F}_L(x_i)\cdot \bm{\mathcal{K}}(x_j,x_i)  \ ,
  \label{sumL}
\end{equation}
where $x_i$ are the nodes of a given quadrature with weights $w_i$. Since the integral in Eq.~(\ref{inner}) is over a charge-conjugate
Bethe-Salpeter amplitude, $\bar \Psi (k,-P) := C\, \Psi^T (-k,-P) C^T$, we introduce in Eqs.~(\ref{sum}) and (\ref{sumL}) ``left'' and ``right'' 
eigenvectors, $\bm{F}_L$ and  $\bm{F}_R$, respectively. Multiplying Eq.~(\ref{sum}) from the left by $\sum_j w_j \, \bm{F}_L'(x_j)$ we obtain, 
\begin{equation}
  \lambda \sum_j  w_j\,  \bm{F}_L' (x_j) \cdot  \, \bm{F}_R(x_j)  
   \ = \ \sum_{i,j} w_i\, w_j \, 
   \bm{F}_L' (x_j) \cdot  \, \bm{\mathcal{K}}(x_j,x_i)\,   \cdot \bm{F}_R (x_i) \  = \  \lambda'\sum_{i} w_i\,  \bm{F}_L' (x_i) \cdot \bm{F}_R (x_i) \ ,
 \label{projecteq}
\end{equation}
where $\bm{F}_L' (x_j)$ is the eigenvector associated with the eigenvalue $\lambda'$.  For $\lambda\ne \lambda^\prime$ this relation implies,
\begin{align}\label{eq:orthogonality}
\int\! dx\ \bm{F}_L' (x) \cdot \bm{F}_R (x)= 0\ , \hspace{0.2cm}
\end{align}
The left eigenvector is given by the Bethe-Salpeter wave function, $S(k-\eta_{-}P)\, \overline{\Phi}_{\lambda'}(k,-P)S(k+\eta_{+}P)$, and the right 
eigenvector is the Bethe-Salpeter amplitude~\cite{Blank:2011qk}. We thus deduce from Eq.~(\ref{eq:orthogonality}),
\begin{align}
  \traceD \int_{k}\, S(k-\eta_{-}P)\,  \overline{\Phi}_{\lambda'}(k,-P)\, S(k+\eta_{+}P)\, \Phi_{\lambda}(k,P)= 0, 
  \hspace{0.2cm}\text{for}\  \lambda'\neq \lambda   \ ,
\end{align}
and because the trace is cyclic this orthogonality condition is equivalent  to that described in the paragraph below Eq.~(\ref{inner}).
Moreover, we verify that the Bethe-Salpeter equation spectrum of Eq.~(\ref{sum}) is equal to that in Eq.~(\ref{sumL}). We stress that 
Eq.~(\ref{eq:orthogonality}) is generally valid and no assumption was made about the kernel structure  or about the eigenvectors.

\section{Examples in hadron physics}
\label{sec5} 

The meson and nucleon resonance structure has been the object of a long history of studies and we here limit ourselves to the approaches based 
on the combined Dyson-Schwinger and Bethe-Salpeter (or Faddeev) equations, in particular their application to excited mesons and nucleons to 
compute their masses, weak decay constants and/or electromagnetic form factors making use of the techniques described in 
Sections~\ref{sec1}--\ref{sec4}. 

Seminal studies on the ground state spectrum of light pseudoscalar mesons established that the $\pi(1300)$ can be described as the first 
radially excited state of the Goldstone boson and furthermore that the decay constant of excited states, $P_n$, vanishes identically in the 
chiral limit~\cite{Holl:2004fr},
\begin{equation}
   f _{P_n}^{\hat m=0} (\mu) \equiv  0  \ ,  \  n \geq  1  \ ,
\label{decayexcited}
\end{equation}
where $\hat m$ is the renormalization-group invariant current-quark mass. Electromagnetic properties of ground and excited state pseudoscalar 
mesons, making use of the the Gram-Schmidt process illustrated in Section~\ref{sec2}, were studied in Ref.~\cite{Holl:2005vu}. 
A first approach to computing the eigenvalues of unflavored light and heavy mesons and their corresponding Bethe-Salpeter amplitudes 
with the implicitly restarted Arnoldi factorization, as implemented in the {\tt ARPACK}~\cite{Arpack} library, is expounded in  
Refs.~\cite{Blank:2010bp,Blank:2011qk} and has successively been applied to the light~\cite{Hilger:2015ora} and heavy~\cite{Hilger:2014nma,Hilger:2015ora} 
quarkonia spectrum for pseudoscalar, vector and tensor states. A recent detailed analysis of the Maris-Tandy interaction~\cite{Maris:1999nt} 
parameter space for ground and radially excited states is presented in Ref.~\cite{Hilger:2015ora}, where the authors conclude that the preferred 
inverse effective range of the Maris-Tandy interaction in rainbow-ladder truncation, $\omega$, is shorter for heavy quarkonia than for the light 
quarkonium spectrum: $\omega_{Q\bar Q} = 0.7$~GeV vs. $\omega_{q\bar q} = 0.5$~GeV.

\begin{table}[t!]
\caption{Masses and decay constants for flavor singlet and nonsinglet $J^{P}=0^-$ mesons; see Section~\ref{sec5}.}
\label{tab1}
\centering
\begin{tabular}{lccl} \hline
                      & Model 1 [GeV]    &   Model 2  [GeV]     &     Reference     \\
\hline
$m_{\pi}$           &  0.138          &       0.153            &  0.139~\cite{Beringer:1900zz}            \\
$f_{\pi}$             &  0.139          &     0.189            &  0.1304~\cite{Beringer:1900zz}           \\ 
\hline
$m_{\pi(1300)}$            & 0.990            & 1.414      &  $1.30\pm 0.10$~\cite{Beringer:1900zz}       \\
$f_{\pi (1300)}$            & $-1.1\times10^{-3}$  & $-8.3\times10^{-4}$  &             \\
       \hline
$m_K$                 & 0.493            & 0.541             &  0.493~\cite{Beringer:1900zz}           \\
$f_K$                 & 0.164            & 0.214              &  0.156~\cite{Beringer:1900zz}            \\
 \hline
$m_{K(1460)}$          & 1.158            &1.580               &   1.460~\cite{Beringer:1900zz}                \\
$f_{K(1460)}$              & $-0.018$        & $-0.017$             &                  \\
  \hline
$m_{\bar ss}$       & 1.287            &1.702               &                 \\
$f_{\bar ss}$       &  $-0.0214$      &  $-0.0216$            &                  \\
  \hline
$m_{\eta_c(1S)}$          & 3.065      & 3.210              &  2.984~\cite{Beringer:1900zz}           \\
$f_{\eta_c(1S)}$      & 0.389            & 0.464             &  0.395~\cite{Davies:2010ip}   \\
 \hline
$m_{\eta_c(2S)}$      & 3.402            & 3.784              &  3.639~\cite{Beringer:1900zz}           \\
$f_{\eta_c(2S)}$      & 0.089           & 0.105              &                  \\
\hline 
\end{tabular}
\vspace{-2mm}
\end{table}

The same library has then been employed to study radial excitations of flavor-singlet and flavored pseudoscalar mesons within the framework of the 
rainbow-ladder truncation~\cite{Rojas:2014aka,El-Bennich:2013yna,Rojas:2014tya}. A summary of the theoretical values for the pseudoscalar's 
ground- and excited-state masses and weak decay constants for flavor-singlet and nonsinglet $J^{P}=0^-$  mesons is reproduced in Table~\ref{tab1}
for two parameter sets of the interaction introduced by Qin et al.~\cite{Qin:2011dd}. Namely, Model~1 and Model~2 correspond to the interaction
parameters, $\omega = 0.4~\text{GeV}$,  $\omega D =(0.8~\text{GeV})^3$ and $\omega =0.6~\text{GeV}$, $\omega D =(1.1~\text{GeV})^3$;
see Ref.~\cite{Rojas:2014aka} for details and Ref.~\cite{Rojas:2014tya} for a discussion of the effective interaction. Note that we explored a
range of parameter combinations for $\omega D \neq$~const., yet could not find a unique set which produces theoretical observables that compare
well with all data in Table~\ref{tab1}. We therefore observe that we are unable to consistently and simultaneously describe both ground and excited 
state observables with the given interaction~\cite{Qin:2011xq} in the rainbow-ladder truncation --- whereas one parameter set successfully reproduces 
experimental ground  state masses and decay constants but not the experimental data on excited states, the second parameter set only provides 
a reasonable description of the excited pseudoscalar mass spectrum.

\begin{figure}[b!]
\vspace*{-8mm}
\includegraphics[width=0.53\textwidth]{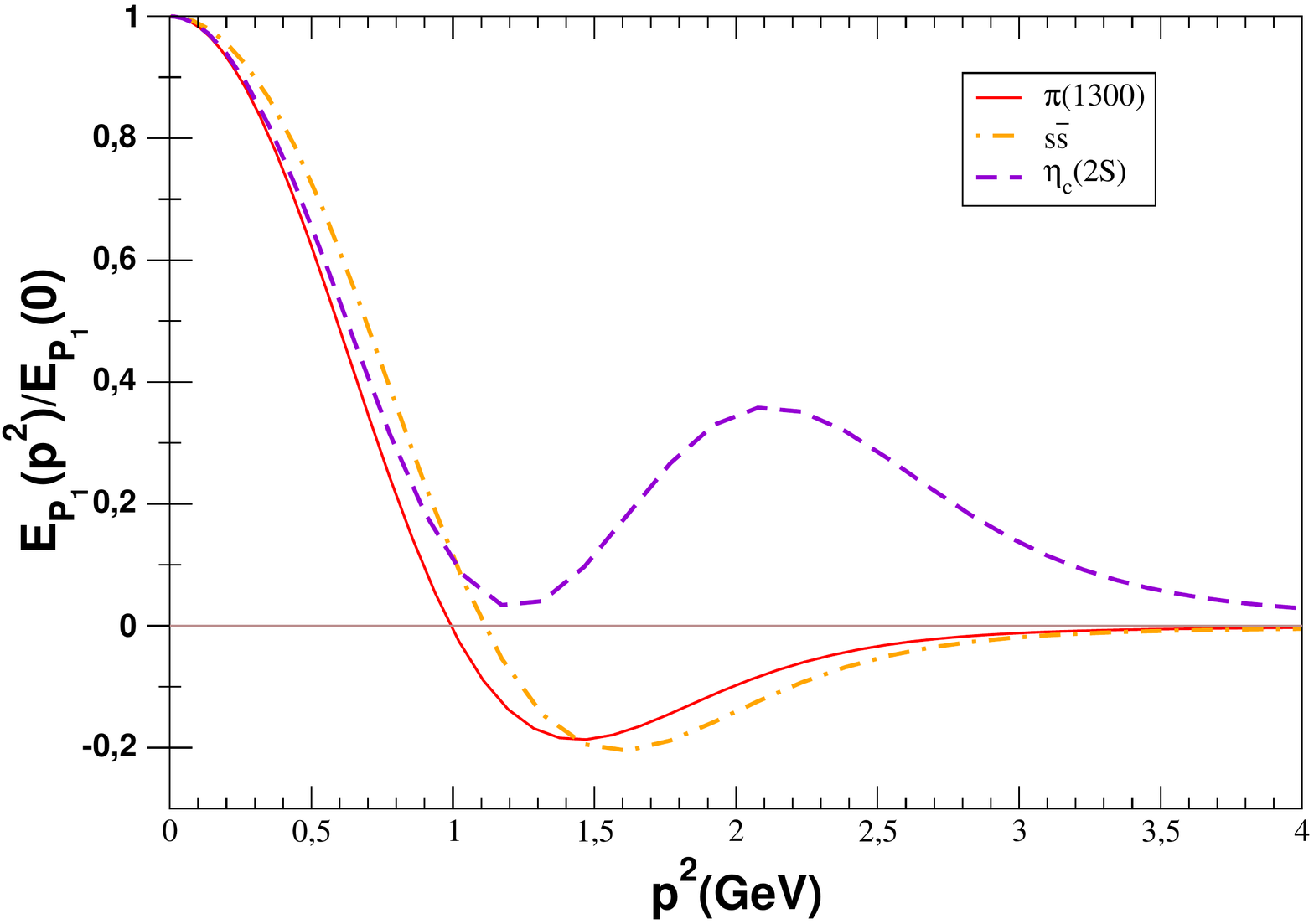}
\includegraphics[width=0.47\textwidth]{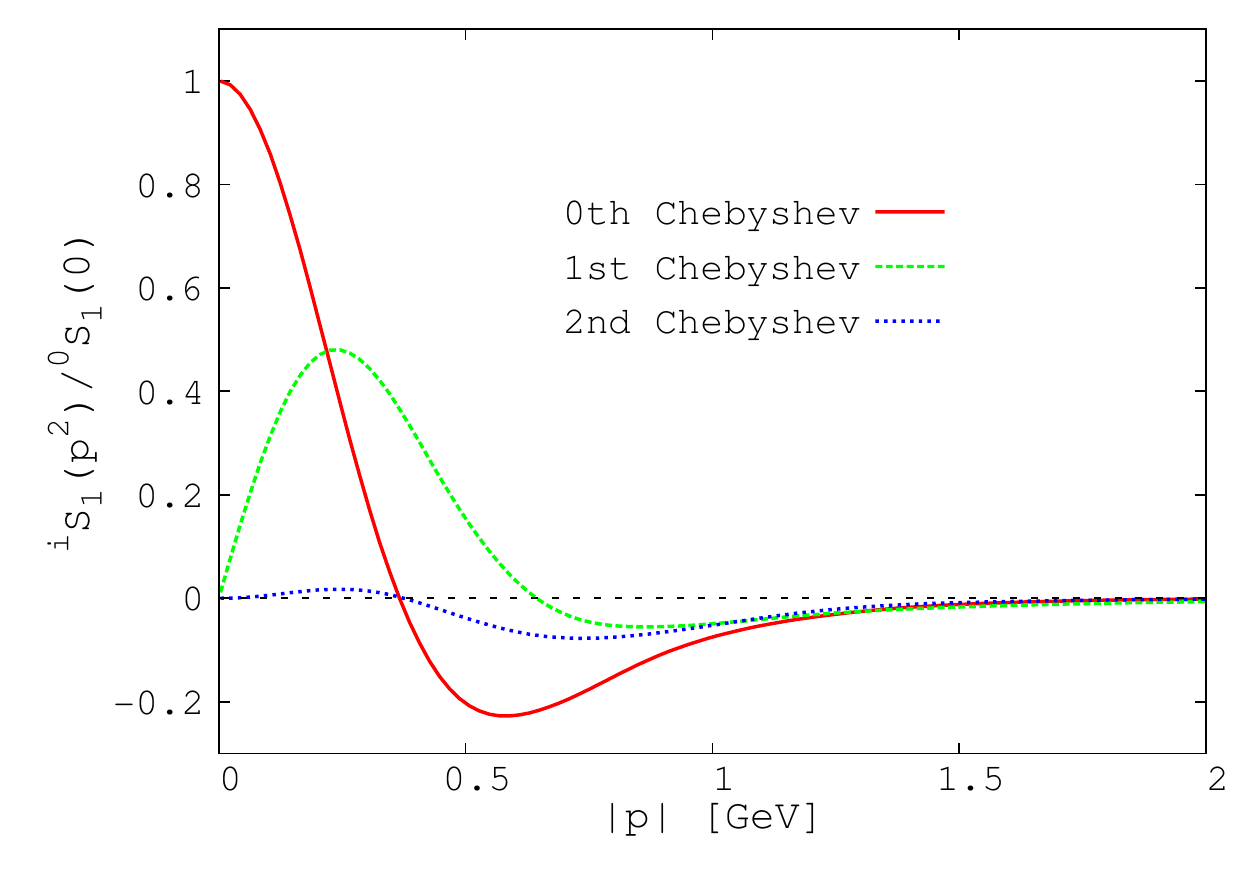}   
\caption{Left panel: Lowest Chebyshev moment, $^0E_{P_1}(p^2)$, associated with the leading Dirac structure $E_{P_1}(p^2)$
   of the pseudoscalar's Bethe-Salpeter amplitude for the first radial excitations $P_1 = \pi(1300)$, $(\bar ss)_{n=1}$ and $\eta_c(2S)$.
   Right panel: First three normalized Chebyshev moments, $^iS_1(p^2)$, $i=0,1,2$, of the leading $S$-wave component in the nucleon's 
   first excited-state Faddeev amplitude.}
\label{fig2}
\end{figure}

As it becomes clear from the left graph of Fig.~\ref{fig2}, for $p^2 \gtrsim 1$~GeV$^2$ the amplitude's lowest Chebyshev projection, 
$^0 E_{P_1}(p^2)$ associated with the leading covariant $i\gamma_5$ of the pseudoscalar's Bethe-Salpeter amplitude, becomes 
negative definite in case of the first radial excitations,  $\pi(1300)$ and $(\bar ss)_{n=1}$, whereas it remains positive for the ground states.
It is remarkable that the behavior of the Bethe-Salpeter amplitudes parallels the pattern of wave functions in quantum mechanics,
namely, the number of zeros can be associated with a principal quantum number $n$. As just mentioned, the ground state amplitude
has no zeros and can thus be associated with $n = 0$. The amplitude of the next highest mass mesons possesses one zero 
and one assigns the quantum number $n=1$ and so on.  An analogous behavior is found for the nucleon's first excited state Faddeev
amplitude for which we plot the first three Chebyshev moments of its leading $S$-wave component. Indeed, very recently 
a range of properties of the proton's radial excitation was predicted which strongly suggests that the nucleon's first radial excitation is the 
Roper resonance~\cite{Segovia:2015hra}. In particular, in Fig.~3 of Ref.~\cite{Segovia:2015hra} one can appreciate that the overlap 
amplitude between the nucleon and its first excited state described by the nucleon-Roper transition form factor $F_1^*(Q^2)$ vanishes 
at $Q^2=0$ and therefore satisfies orthogonality.

\begin{acknowledgement}
We thank the organizers of the {\em 21 International Conference on Few-Body Problems in Physics\/} in Chicago for the opportunity 
to present our work at this successful event. The work of B.~E. is supported in parts by a State of S\~ao Paulo Research 
Foundation (FAPESP) grant, an MCTI/CNPq/Universal grant and by the Brazilian agency CNPq with a visiting fellowship at 
the Institute for Theoretical Physics, State University of S\~ao Paulo (IFT-Unesp). The work of E.~R. is supported  by 
``{\em Patrimonio Aut\'onomo Fondo Nacional de Financiamiento para la Ciencia, la Tecnolog\'ia y la Innovaci\'on, Francisco 
Jos\'e de Caldas}'' and by ``{\em Sostenibilidad-UDEA 2014-2015}''.
\end{acknowledgement}

%

\end{document}